\documentclass[twocolumn,preprintnumbers,superscriptaddress]{revtex4-2}
\usepackage{physics}
\usepackage{utfsym}

\usepackage{hyperref}
\usepackage{graphicx}
\usepackage{amsmath}
\usepackage{amssymb}
\usepackage[OT1]{fontenc}
\usepackage{cmbright}
\DeclareFontShape{OT1}{cmss}{m}{it}{<->ssub*cmss/m/sl}{}

\usepackage{listings}
\usepackage{xcolor}
\usepackage{braket}
\usepackage{xspace}
\usepackage{makecell}
\usepackage[capitalise, noabbrev]{cleveref}

\begin{document}

\title{\Qibosoq: an open-source framework for quantum circuit RFSoC programming}

\preprint{TIF-UNIMI-2023-18, CERN-TH-2023-180}

\newcommand{\MIaff}{TIF Lab, Dipartimento di Fisica, Universit\`a degli Studi
	di Milano and INFN Sezione di Milano, Milan, Italy.}

\newcommand{\TII}{Quantum Research Center, Technology Innovation Institute, Abu Dhabi, UAE.}

\newcommand{\UNIMIB}{Dipartimento di Fisica, Universit\`a di Milano-Bicocca, I-20126 Milano, Italy.}

\newcommand{\INFNUNIMIB}{INFN - Sezione di Milano Bicocca, I-20126 Milano, Italy.}

\newcommand{\CERNaff}{CERN, Theoretical Physics Department, CH-1211
	Geneva 23, Switzerland.}

\newcommand{\UABaff}{Departament de Telecomunicació i Enginyeria de Sistemes, Universitat Autonoma de Barcelona, ES-08193 Barcelona, Spain}

\newcommand{\BIQUTE}{Bicocca Quantum Technologies (BiQuTe) Centre, I-20126 Milano, Italy.}

\newcommand{\Fermilab}{Fermi National Accelerator Laboratory, PO Box 500, Batavia IL, 60510, USA.}

\newcommand{\Princeton}{Department of Physics, Princeton University, Princenton, New Jersey, USA.}

\author{Rodolfo Carobene}
\email{rodolfo.carobene@mib.infn.it}
\affiliation{\UNIMIB}
\affiliation{\INFNUNIMIB}
\affiliation{\TII}
\author{Alessandro Candido}
\affiliation{\MIaff}
\affiliation{\CERNaff}
\author{Javier Serrano}
\affiliation{\TII}
\affiliation{\UABaff}
\author{Alvaro Orgaz-Fuertes}
\affiliation{\TII}
\author{Andrea Giachero}
\affiliation{\UNIMIB}
\affiliation{\INFNUNIMIB}
\affiliation{\BIQUTE}
\author{Stefano Carrazza}
\affiliation{\CERNaff}
\affiliation{\MIaff}
\affiliation{\TII}

\newcommand{\Qick}{\texttt{Qick}\xspace}
\newcommand{\Qibosoq}{\texttt{Qibosoq}\xspace}
\newcommand{\Qibo}{\texttt{Qibo}\xspace}
\newcommand{\Qibocal}{\texttt{Qibocal}\xspace}
\newcommand{\Qibolab}{\texttt{Qibolab}\xspace}

\begin{abstract}
	We present \Qibosoq, an open-source server-side software package designed for
	radio frequency system on chip (RFSoC) for executing arbitrary pulse sequences
	and algorithms on self-hosted quantum processing units using only open-source software.
	\Qibosoq connects the RFSoC firmware provided by \Qick, a Quantum
	Instrumentation Control Kit, with \Qibo, a quantum computing middleware
	framework that enables both experimental and gate-based applications.
	It simplifies the work of experimentalists and developers by managing client-server communication protocols, implementing tests, and validation procedures, thereby reducing the complexity of experimental setups.
	The client-side integration is achieved with dedicated
	drivers implemented in \Qibolab, the specialized software module of \Qibo for
	quantum hardware control. Therefore, this setup provides a seamless mechanism
	to deploy circuit-based algorithms on custom self-hosted quantum hardware
	platforms controlled by RFSoC electronics.
	%
\end{abstract}

\maketitle

\tableofcontents

\section{Introduction}

Controlling superconducting qubits and other quantum technologies necessitates instruments and software drivers capable of generating and modulating arbitrary pulses in the microwave radio frequency range~\cite{Riste2020}.
With qubits being extremely delicate systems, reliable and efficient control
electronics are mandatory for the successful operation of quantum hardware~\cite{Ladd2010}.

Nowadays, one of the challenges of research institutions is to identify
the proper set of instruments with the desired specifications
for quantum technologies. An important figure of merit, other than
performances, is price. In the last decade, new proprietary commercial
products dedicated to quantum computing have been released, some examples
include Qblox~\cite{Qblox}, Quantum Machines~\cite{quantum_machines}, Zurich
Instruments~\cite{zurich_instruments} among others. Despite the interest from
manufacturers in commercializing expensive ready-to-run solutions, experimental
laboratories still have to face the issue of acquiring instruments which are in
continuous improvement in terms of firmware and software, therefore customers
might participate indirectly as co-developer by providing feedback, testing and
waiting for improvements.
This situation is not ideal as it increases the manpower and time required by research teams in order to define the experimental protocols.
\Qibosoq addresses this by providing an open-source alternative that is intigrated in a full-stack framework.

Radio Frequency System on Chip~\cite{Javaid2022,Gebauer2021,Tholen2022,Singhal2023} (RFSoC)
FPGA (Field Programmable
Gate Arrays) is a low-cost hardware alternative which provides flexible
development of firmware and software related to quantum technologies. The research
community has already achieved open-firmware for quantum applications through
the \Qick (Quantum Instrument Control Kit) project~\cite{Stefanazzi2022},
that enables the use of RFSoCs to generate the sequences of pulses required for
controlling and reading out qubits. Moreover, \Qick also provides a higher-level Python
library that eases the execution of pulses.

\begin{figure}
	\includegraphics[width=0.4\columnwidth]{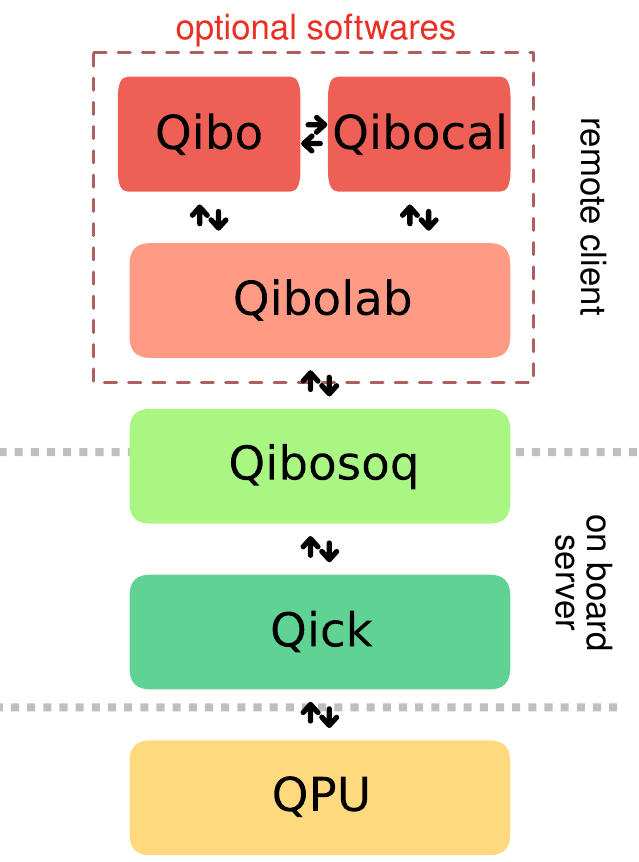}
	\caption{\label{fig:qibo-qibosoq-qick}Deployment pipeline from \Qibo to the
		quantum processing unit (QPU). \Qibosoq can also be used as a standalone server
		library, the objects in the dashed area are optional and only required to
		connect \Qick and \Qibosoq with \Qibo.}
\end{figure}

In this manuscript, we present for the first time \Qibosoq~\cite{rodolfo_carobene_2023_8223402}, an open-source
software package which unlocks \Qick's potential to execute quantum
algorithms on self-hosted quantum hardware platforms through \Qibo, a quantum
computing
framework~\cite{Efthymiou_2021,Efthymiou_2022,Carrazza_2023,stavros_efthymiou_2023_7736837}
and its hardware module \Qibolab~\cite{Efthymiou_2023}.
\Qibosoq provides a dedicated application programming interface (API)
for sending arbitrary pulses sequences to the \Qick firmware.
It is composed both of a server component, running on the RFSoC along with \Qick, and of a remote client
that can be attached to \Qibolab or potentially other higher-level softwares.
In \cref{fig:qibo-qibosoq-qick} we show a schematic view of the software stack
from the quantum processing unit (QPU) to \Qibo. The full pipeline is required
for the execution of quantum circuit algorithms, but the user can also use directly
\Qibolab for setting up experiments related to pulse generation.
The stack is designed in a modular way, therefore \Qibosoq can also be used
as a standalone server application which runs on RFSoC FPGA boards
opening the possibility to interface \Qick with multiple experimental configurations.

Having the \Qick firmware integrared into the \Qibo framework also unlocks the
possibility of writing automated calibration and recalibration experiment without
the need to define every protocol from scratch (leveraging
\Qibocal~\cite{andrea_pasquale_2023_7957542}).
Moreover, various algorithms are already defined in \Qibo (for example variational
quantum algorithms for chemistry applications and error mitigation schemes for
arbitrary circuits), so the integration provided by \Qibosoq allows any laboratory
with superconducting qubits and a supported RFSoc, to control the full
Quantum-computing stack with the same integrated framework.

The paper's structure is as follows. In \cref{sec:methodology} we describe a
detailed overview of the \Qibosoq library for version \texttt{0.1.3}. In
\cref{sec:results} we show performance benchmark results and example results
of calibration experiments. Concluding our study, we provide a summary and
outline future development directions in Section \ref{sec:outlook}.

\section{Methodology}
\label{sec:methodology}

In this section, we will provide a brief overview of the current status of
the \Qick project. We will then delve into the specifics of \Qibosoq, including
its internal layout and the comprehensive set of supported features.

\subsection{The QICK project}\label{sec:qick}

Introduced in 2022, the \Qick project offers an open-source qubit controller
based on the Zynq UltraScale+ Xilinx ZCU111~\cite{xilinxZCU111} evaluation board,
supporting direct radio frequency (RF) synthesis up to approximately $6$ GHz via
the XCZU28DR RFSoC.

It consists of a custom FPGA firmware and a Python library for coding experiments.

The firmware leverages the RFSoC's digital-to-analog converters (DACs) as arbitrary
waveform generators for RF pulses and enables fast, precise acquisitions through
the analog-to-digital converters (ADCs). Additionally, it features a custom
timed-processor (tProcessor) that allows users to program sequences of timed
pulses and loops using an assembly-like language.

Since its initial presentation, the \Qick firmware has undergone several updates
and now supports the evaluation boards RFSoC4x2~\cite{xilinxRFSoC4x2} and
ZCU216~\cite{xilinxZCU216} as well, mounting respectively the XCZU48DR and
XCZU49DR chip. New firmware versions have been developed with support for
multiplexing, enabling readouts on different qubits coupled to the same line
using the same DAC and ADC. Moreover, the team is still exploring new functionalities,
including resonator simulators on the FPGA, constant outputs and more specific DAC
configurations.

The role of the software side of \Qick is to eliminate the need for writing
experiments directly in the tProcessor language. Instead, it allows
users to write experiments in the Python  programming language. It accomplishes
this by offering helper functions and  program templates that streamline the
coding experience for users, while still leveraging the capabilities of the tProcessor.

\Qick is currently used in different labs around the world, with application in different
technologies such as superconducting qubit control, entangled photon pair generation, and
SNSPD (Superconducting Nanowire Single-Photon Detectors) readout.
Several experiments have been already published using the \Qick to control superconducting
quantum hardware~\cite{Bryon2023,Martinez2023,Xie2023,Anferov2023,Efthymiou_2023,Moskalenko2022}.

The introduction of \Qick has already proven to be beneficial for researchers
in developing custom solutions for qubit control.
However, from a usability perspective, there are still some limitations.
For example, to fully utilize the tProcessor's functionalities, users are required
to write more advanced code, even when using the Python API.
This entails managing memory, registers, delays related to the processor latency
and not only to the experiment itself, as well as other low-level aspects which
can be cumbersome and error-prone for average users.
Moreover, \Qick operates exclusively at a pulse-level abstraction, and while
this makes it a powerful tool for controlling quantum systems and executing
pulse-based experiments, it is not directly suitable for quantum computing
applications that rely on gate-based circuits.

In this context, \Qibosoq plays an important role by leveraging the existing \Qick
software and firmware to simplify the coding of quantum computing experiments in
the tProcessor language. It offers a more user-friendly interface, allowing researchers
to focus on their experiments, abstracting away the low-level details of the tProcessor.
The experiments defined are still just based on pulses, but since \Qibosoq is natively
integrated with the \Qibo framework, performing circuit-based experiments
become straightforward through \Qibolab. Moreover, with this integration \Qick
becomes also connected with \Qibocal~\cite{andrea_pasquale_2023_7957542} that
enables simplified and automated qubit calibration. Lastly, it is possible to
use directly \Qibo to execute algorithms, that are first converted into pulses
by \Qibolab and then executed on the RFSoC using \Qibosoq.
Overall, the connection between \Qibosoq and \Qibo extends the capabilities of \Qick,
enabling users to work at both the pulse-level with additional tools and
gate-level abstractions.
We believe that researchers involved with studying and calibrating qubits (or
in general superconducting devices) could be interested in using \Qibosoq in
connection with \Qibolab and \Qibocal in order to focus on writing new experiments,
without the need to re-write already existing ones.
On the other hand, even researchers only interested in executing gate-based
algorithms, without any specific knowledge on the control and readout of a
superconducting qubit, could leverage the integration between \Qibosoq and
\Qibo in order to perform experiments on self-hosted qubits, while completely
ignoring the gates implementation.

\subsection{Networking}

The boards supported by \Qick share a common architecture, that includes an
on-board CPU, beyond the actual FPGA device. This design makes them
independent, not requiring a further processor to execute applications.
\Qick itself is organized to run on this embedded processors, based on
\textit{Pynq}~\cite{Xilinx2018} structure.

However, the typical QPU user will not have direct access to the board, writing
or loading its software there, but the access will be granted through a
network.
For this reason, since \Qibosoq is already operating at a higher level than
\Qick, it is worth to take into account the networking part, controlling the
communication between the board and the end user, making it as simple as
possible, while flexible enough to support many different kinds of
applications.

The communication layer can happen at different levels, e.g.\ it could transfer
entire executables and run them on-board or transmit just minimal set of instructions.
This might be advantageous to allow lower level access to the user, thus being
free to access primitives from the programming language and \Qick library.
But this is not the typical use case, since most of the QPU operations share a
common layout.

Recognizing the common elements allows us to factorize them out from the
multiple applications into a single library, reducing duplication and
simplifying the applications development.
Furthermore, restricting the applications' degrees of freedom reduces the
amount of information that has to be transferred on the network, eventually
providing a higher communication efficiency.
These considerations led the design of \Qibosoq  and
its own internal \textit{language} for experiments description.

In this matter, there are other two considerations that is worth taking into account.
One is related to the available on-board CPUs, the other to a partial current \Qick
limitation.

When taking into account where to run the application bulk, the nature of the
application itself has to be examined.
Some applications require to run just QPU experiments and collect their
data, for further elaboration later on. This kind of application is QPU-bound,
and it has very limited requirements in terms of classical processing.
However, more hybrid applications are also possible: a typical example being a
Quantum Machine Learning (QML) with classical optimization, where the QPU usage
is interleaved with classical computation.
When coming to hybrid applications there are two competing factors: latency and
performances. Limiting possible applications to run on the on-board CPUs it
would not scale, i.e. it would not allow to keep good performances for
increasingly larger problems (as compared to the time required for execution on
other commonly available hardware), becoming soon the main bottleneck.
On the other hand, off-loading the computation to a separate machine introduces
communication latency during the application execution.
The chosen trade-off in \Qibosoq values more the potential application scaling,
giving it the flexibility to support arbitrary kinds of hybrid applications, at
the expense of time performances. This takes into account the current limited
performances of the on-board CPUs. For instance, the Xilinx~\cite{xilinx}
boards feature ARM Cortex-A53 processors, which are considerably slower even
than processors found in most modern laptops.
In future, different type of boards might become available, and they could be
developed having in mind quantum hybrid applications. Or the communication
with an external processors might improve, not relying on a LAN but making use
of local buses, like Peripheral Component Interconnect (PCI).
\Qibosoq approach is easily extendable to this improved scenarios, in which the
same (or similar) internal language could be maintained, just acting on the
communication layer implementation.

Instead, the second consideration is purely technical: each experiment in \Qick
requires a certain common initialization. Repeating this initialization has two
main drawbacks, since on the one side it requires to pay every time the
performance cost, and on the other it resets the ADCs and DACs clocks, thereby
introducing a random phase between them, and consequently loosing phase
coherence between experiments.
This phase coherence is critical in quantum computing applications as it allows
for qubit calibration to be maintained between executions, eliminating the need
of a partial recalibrating every run.
Because of this, it is necessary to keep a \Qick instance alive between
multiple user connections. This limitation requires to have at least a minimal
server running on board.
While \Qick already provides a way of solving this problem, leveraging the \textit{Pyro4}
library that is used to send the required objects through a network, it is still
designed as an on-board software and the integration is not straightforward.
\Qibosoq is built as an alternative to this solution to make it easier to control
\Qick remotely, while also simplifying its control and integrating it with \Qibo.
In this sense, \Qibosoq is just an extension of \Qick, offering a higher level
interface, preferable for the end user applications development.

It is relevant to note that, despite having discussed of \textit{user
	applications} until here, a \Qibosoq user might also be a higher level library,
dealing with some kind of task involving QPU execution. \Qibolab and \Qibocal
are two examples of this layout, and they showcase the flexibility of \Qibosoq,
to support arbitrary execution through an RFSoC controller.
In particular, \Qibolab also allows the execution of arbitrary circuits,
described in the \Qibo language, and shows how it is possible to deal with the
transpilation and experiment preparation on the \Qibosoq client side, making
use of arbitrary classical resources.

Instead, \Qibocal is a perfect example of
how multiple clients can share the same server, since its calibration routines
are all potentially independent, and being executed as fully separate programs.
The only requirement to interface with the \Qibosoq server is to describe the
final QPU execution using \Qibosoq primitives, and finally establish a
connection to the server.

\subsection{Qibosoq layout}

\Qibosoq is composed of multiple submodules: the \textbf{programs}, representing
\Qick programs, and eventually taking care of the experiment compilation into
the tProcessor language, the \textbf{components}, high-level structures used
in the \textit{programs} construction, the \textbf{server} implementation, and
\textbf{client} utilities, to manage the communication.

\begin{figure}
	\includegraphics[width=0.8\columnwidth]{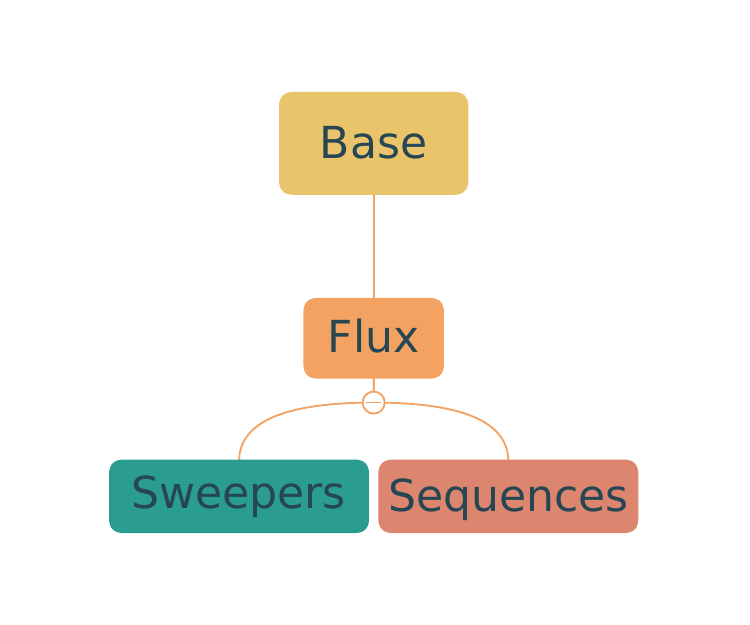}
	\caption{Hierarchy in the \textbf{programs} submodule}
	\label{fig:programs}
\end{figure}

The \textbf{programs} bridge the gap between the high-level interfaces (components)
and the low-level execution on quantum hardware.
Eventually, only two distinct programs are directly used, but the full hierarchy
(presented in \cref{fig:programs}) also includes intermediate abstractions.
Considering all layers, the defined \textbf{programs} are:
\begin{itemize}
	\item the abstract \textbf{base} program, that contains functions shared among all possible experiments and executions. It serves as the foundation for all the other \Qibosoq programs;
	\item the abstract \textbf{flux} program, that collects the additional elements required for controlling flux-tunable qubits. In addition to the functionalities defined in \textbf{base}, it includes support for bias voltages and fast DC (direct current) pulses;
	\item the \textbf{sequences} and \textbf{sweepers} programs that contain the different elements used, respectively, in the execution of fixed parameters pulse sequences and real-time sweeps. They inherit all the functionalities defined in \textbf{base} and \textbf{flux}, while also being specific implementations of \Qick classes: the \texttt{AveragerProgram} and the \texttt{NDAveragerProgram}.
\end{itemize}

With this set of \textbf{programs} it is possible to define a large variety
of experiments. Indeed, the number of drive pulses, of readouts and flux pulses, is
only limited by the on-board available memory. And various acquisition modes are
available for all the combinations of pulses.
Moreover, using \textbf{sweepers} it is possible to speed up real experiments,
taking the best out of the tProcessor speed.

There is no requirement for the final user to have direct knowledge of which program subclass is used, since \Qibosoq takes care of it automatically.

The \textbf{components} play the crucial role of establishing a common language
for communication, easing the implementation of a \Qibosoq client in \Qibolab
or by other parties.
The main elements defined within the \textbf{components} submodule include:
\begin{itemize}
	\item the \textbf{Config} object that contains essential general information
	      required for executions. This includes the number of software and hardware
	      repetitions, delay between repetitions (to ensure qubit relaxation), and
	      whether to average the results among repetitions or not;
	\item the \textbf{Pulse} base object that serves as the foundation for different
	      implemented pulse shapes. Rectangular, Gaussian and DRAG~\cite{Gambetta2011}
	      pulses are natively supported, as well as custom waveform shapes defined by
	      their ''in-phase`` and ''quadrature`` (IQ)~\cite{Franks1969,Naidu2003} values;
	\item the \textbf{Qubit} object that holds information about any necessary bias
	      required for operating it;
	\item the \textbf{Sweeper} and \textbf{Parameter} objects that are used to
	      describe real-time on-hardware scans.
\end{itemize}

Using these components, the typical \Qibosoq program will start with something similar to
\begin{lstlisting}
from qibosoq.components.base import (
    Qubit,
    Config
)
from qibosoq.components.pulses import Rectangular

# Definition of a pulse
pulse = Rectangular(
          frequency = 5000,    # MHz
          amplitude = 1,       # FullScale
          relative_phase = 0,  # Rad
          start_delay = 1,     # us
          duration = 1,        # us
          name = "mypulse",
          type = "drive",
          dac = 1,
)

# Definition of the sequence
sequence = [pulse, pulse, pulse]

# General configuration of the experiment
config = Config(
    relaxation_time = 50,
    reps = 10000,
    average = False
)

# Definition of biases
qubit = Qubit(bias=0.1, dac=2)
\end{lstlisting}
Note that almost every line of this snippet was related to experiment parameters.

\texttt{Pulse}, \texttt{Sequence}, \texttt{Config} and \texttt{Qubit} are all
mandatory objects that are used in any experiment, while \texttt{Sweeper} objects
need to be defined only when used.

The last two fundamental elements in \Qibosoq, are the \textbf{client} and
the \textbf{server}. The \textbf{client} is composed of a set of tools used
to connect to the server, convert components into a serialized form, and send
them following the \Qibosoq communication protocol. The \textbf{server}
implements the on-board server, continuously listening for connections, and
executing received instructions by initializing and running the required
programs on the quantum hardware.

Albeit these are fundamental parts of \Qibosoq, the user usually will not directly
care about them, interacting only with the more user-friendly interface. Only on
startup of the board it is required to start the server with the command-line:
\begin{lstlisting}[language=bash]
python -m qibosoq
\end{lstlisting}
A full description of the API and all Qibosoq capabilities can be found in the
online documentation~\cite{qibosoq_doc}, alongside extended tutorials teaching
how to use it to implement various experiments.

\subsection{Serialization and communication protocol}\label{sec:communication}

To run an experiment, users in the client need to define instructions using
\Qibosoq components, describing the experiment to be executed.
These instructions are then sent to the server using the Transmission Control
Protocol (TCP) for communication. TCP was chosen for its reliability~\cite{Kanmai2020},
since it ensures data integrity, preventing any loss during transmission.
Additionally, its session-based protocol suits \Qibosoq's requirements by
disallowing multiple concurrent executions, ensuring smooth and well-controlled
communication between the server and potentially multiple clients.

While the server is composed only of \Qibosoq code, the client can be part of
very different frameworks or even be just a standalone script, as long as it
sends to the server the expected commands in the expected format, \Qibosoq will
work properly. Through the \textbf{client} module, \Qibosoq offers some helpers
to make following the communication protocol easier.

\begin{figure}
	\includegraphics[width=0.8\columnwidth]{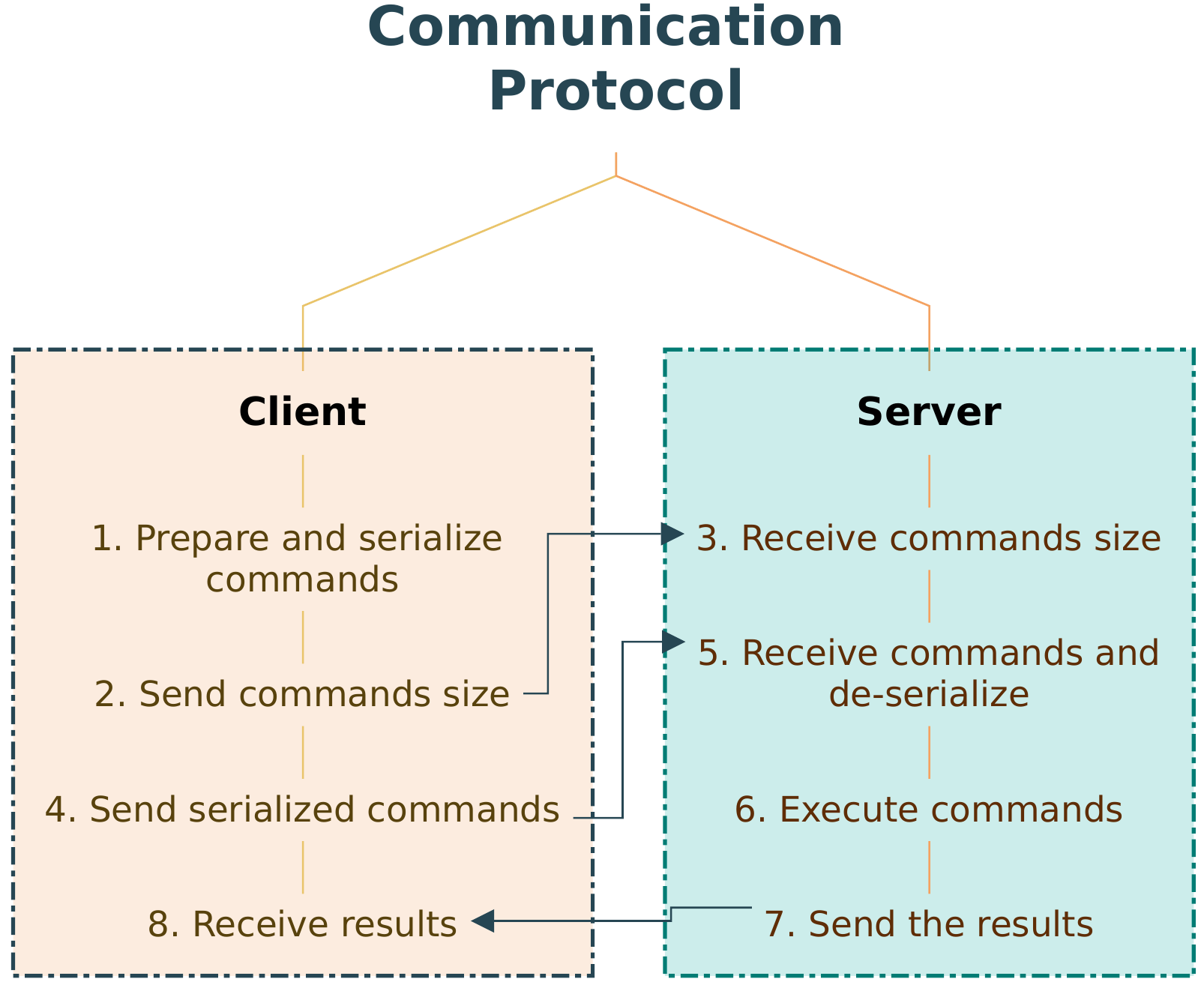}
	\caption{Schematic of the communication protocol.}
	\label{fig:communication}
\end{figure}

The \Qibosoq communication protocol is schematically presented in
\cref{fig:communication}.

The first step in the process is defining the experiment to be executed,
utilizing the \Qibosoq components:
\begin{description}
	\item[operation] \Qibosoq supports different execution modes that can be
	      selected through a specific component. These modes enable the execution
	      of different type of experiments:
	      \begin{itemize}
		      \item fixed-parameters experiments defined as a sequence of pulses.
		      \item varying-parameters experiments defined using a combination
		            of sequence of pulses and sweepers.
		      \item raw acquisition experiments, where the raw signal (after
		            demodulation) is acquired without integration.
	      \end{itemize}
	\item[configuration] a \textbf{Config} object with general experiment parameters;
	\item[sequence] a list of pulses that describe the experiment, including control,
	      flux, and readout pulses for performing measurements. Each pulse contains
	      information about its shape, frequency, and start time.
	\item[qubits] a list of qubit objects.
	\item[sweepers] a list of sweepers, only required for real-time parameter scans
	      on the FPGA logic. The sweepers can act on multiple parameters simultaneously
	      or update them sequentially, enabling exploration of all combinations of chosen
	      parameters during the experiment.
\end{description}

This set of instruction provides a comprehensive description of the experiment
to be executed and offer a flexibility that can accommodate various qubit-related
experiments.

The first step ends with the serialization, where the instructions are first dumped
in the JSON format and then encoded into bytes using UTF-8.

Then, a connection is established between the client and server.
Following the TCP three way handshake, a first packet containing a 32-bit integer
(four bytes) is sent. This integer represents the byte-size of the instructions
to be transmitted immediately afterward.

Upon receiving and de-serializing the commands, the server initializes the required
program based on the specified \textbf{operation} parameter. After executing the
experiment, the server sends back the results, comprising acquired ``i'' and ``q''
values, through the same TCP connection to the client.

Once the client receives all the data, the connection is closed, and the server
returns to waiting for new commands.

These operations are all executed automatically by the \Qibosoq package, so that
the user does not have to take care of almost anything that is not related to the
experiment itself. After definition of the sequence and required objects, the user
is only required to write the commands in a dictionary for execution:
\begin{lstlisting}
from qibosoq.client import execute

server_commands = {
    "operation_code": OperationCode.EXECUTE_PULSE_SEQUENCE,
    "cfg": config,
    "sequence": sequence,
    "qubits": [qubit],
}

# Perform the experiment
HOST = "192.168.0.0"  # Server IP address
PORT = 6000 # Default server port
i, q = execute(server_commands, HOST, PORT)
\end{lstlisting}

Note that, while \Qibosoq already provides a client that takes care of implementing
this communication protocol, it does not strictly require its use, and potentially
other clients could implement the same protocol and interact with the \Qibosoq
server part.

\subsection{Features and limitations}

\Qibosoq abstracts a higher-level interface over the \Qick primitives, and its
current main purpose has been to serve \Qibolab, despite not being restricted to it.
Therefore, it is relevant to be aware of the present limitations of these related libraries when approaching \Qibosoq, and how \Qibosoq itself is affected by them.

In \cref{tab:supported-features} a small comparison of the main features supported by \Qick, \Qibosoq and \Qibolab are presented.

\begin{table}
	\begin{tabular}{lccc}
		\hline \hline
		\textbf{Feature}          & \textbf{Qick}                                                                                               & \textbf{Qibolab} & \textbf{Qibosoq} \\ \hline
		Arbitrary pulse sequences & \usym{1F5F8}                                                                                                & \usym{1F5F8}     & \usym{1F5F8}     \\
		Arbitrary waveforms       & \usym{1F5F8}                                                                                                & \usym{1F5F8}     & \usym{1F5F8}     \\
		Multiplex readout         & \usym{1F5F8}\footnote{\label{footnote:special_firmware}Special firmware available from \Qick under request} & \usym{1F5F8}     & \usym{1F5F8}     \\
		Feedback                  & \usym{1F5F8}                                                                                                & \usym{1F5F8}     & \usym{26ED}      \\
		RTS frequency drive       & \usym{1F5F8}                                                                                                & \usym{1F5F8}     & \usym{1F5F8}     \\
		RTS frequency readout     & \usym{2613}\footref{footnote:special_firmware}                                                              & \usym{1F5F8}     & \usym{2613}      \\
		RTS amplitude             & \usym{1F5F8}                                                                                                & \usym{1F5F8}     & \usym{1F5F8}     \\
		RTS duration              & \usym{2613}\footnote{Supported for specific pulse shapes}                                                   & \usym{1F5F8}     & \usym{2613}      \\
		RTS start                 & \usym{1F5F8}                                                                                                & \usym{1F5F8}     & \usym{1F5F8}     \\
		RTS relative phase        & \usym{1F5F8}                                                                                                & \usym{1F5F8}     & \usym{1F5F8}     \\
		RTS N-Dimensional         & \usym{1F5F8}                                                                                                & \usym{1F5F8}     & \usym{1F5F8}     \\
		Hardware averaging        & \usym{1F5F8}                                                                                                & \usym{1F5F8}     & \usym{1F5F8}     \\
		Singleshot (No Averaging) & \usym{1F5F8}                                                                                                & \usym{1F5F8}     & \usym{1F5F8}     \\
		Integrated acquisition    & \usym{1F5F8}                                                                                                & \usym{1F5F8}     & \usym{1F5F8}     \\
		Classified acquisition    & \usym{1F5F8}                                                                                                & \usym{1F5F8}     & \usym{1F5F8}     \\
		Raw waveform acquisition  & \usym{1F5F8}                                                                                                & \usym{1F5F8}     & \usym{1F5F8}     \\
		\hline \hline
	\end{tabular}
	\caption{Main features and limitations of \Qick, \Qibosoq and \Qibolab compared.
		The features denoted by `` \protect\usym{1F5F8} '' are supported, `` \protect\usym{2613} ''
		means not supported and `` \protect\usym{26ED} '' under development.}
	\label{tab:supported-features}
\end{table}

Here, "RTS`` refers to the term "Real Time Sweeper``, denoting the ability to conduct
parameter scans directly on FPGA logic. For a comprehensive explanation of the mentioned
features, please consult~\cite[Section~III.B]{Efthymiou_2023}.

This selection of features, commonly used in standard qubit experiments, are
all supported by \Qibosoq, with the notable exception of the feedback feature.

Using \Qibolab, is also possible to access all of these features with the sole
exception of sweepers on multiple different parameters with concurrent updates,
that is also currently under development.

Some more limitations come directly from the hardware.
\Qibosoq is compatible with all the boards supported by \Qick, namely the RFSoC4x2,
the ZCU216 and the ZCU111.
In \cref{tab:supported-boards}, the three RFSoCs are presented along with some details.

Note that, to control flux-tunable qubits, the companion boards included in the
standard Xilinx kits are not sufficient,
since they usually include also baluns on the single-ended outputs/inputs
connected to the DACs and ADCs that filter the DC currents required for controlling
the qubits. This problem can be solved by using the available differential outputs
along with some differential amplifiers (as Texas Instruments
THS3217~\cite{texasinstruments_THS3217}) to convert the signal from double to
single-ended or with the use of custom companion boards~\cite{Stefanazzi2022,Tholen2022}.

\begin{table}
	\begin{tabular}{lccc}
		\hline \hline
		                                                                                                                                                          & \textbf{ZCU111} & \textbf{RFSoC4x2} & \textbf{ZCU216} \\ \hline
		Physical DACs                                                                                                                                             & 8               & 2                 & 16              \\
		\Qick activated DACs\footnote{\label{footnote:firmware}These numbers are related to the standard available firmware, they can vary with other firmwares.} & 7               & 2                 & 7               \\
		DAC sampling rate [GSPS]                                                                                                                                  & 6.554           & 9.85              & 9.85            \\
		Physical ADCs                                                                                                                                             & 8               & 4                 & 16              \\
		\Qick activated ADCs\footref{footnote:firmware}                                                                                                           & 2               & 2                 & 2               \\
		ADC sampling rate [GSPS]                                                                                                                                  & 4.096           & 5                 & 2.5             \\
		\hline \hline
	\end{tabular}
	\caption{Outline of the RFSoCs supported by \Qibosoq and their characteristics.}
	\label{tab:supported-boards}
\end{table}

\section{Results}
\label{sec:results}

All the experiments presented in this section were performed using \Qibocal,
the runcards required to reproduce them are openly available at~\cite{qibosoq_runcards}.

\begin{figure*}
	\includegraphics[width=1\textwidth]{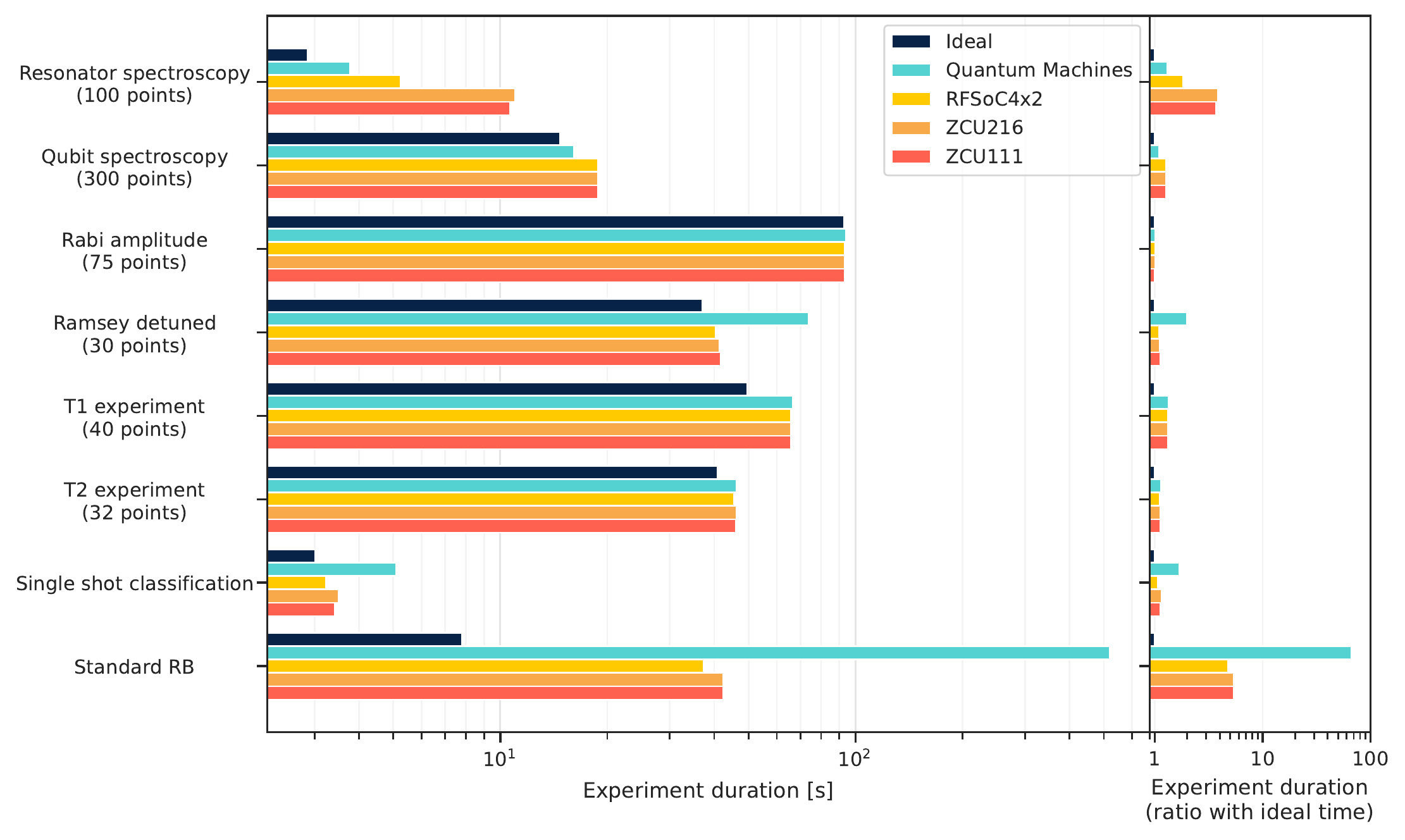}
	\caption{Execution time of various qubit calibration routines on different electronics. On the left side we show the absolute times in seconds for each experiment. The black bar, that represents the ideal time, shows the minimum time theoretically required for each experiment. On the right side we show the ratio between actual execution time and ideal time, the difference from ideal and real time comes from software delays and communication latency.}
	\label{fig:speed_benchmark}
\end{figure*}

\subsection{Cross-platform benchmark}
\label{sec:benchmark}

\begin{figure*}
	\begin{tabular}{ccc}
		\includegraphics[width=0.33\textwidth]{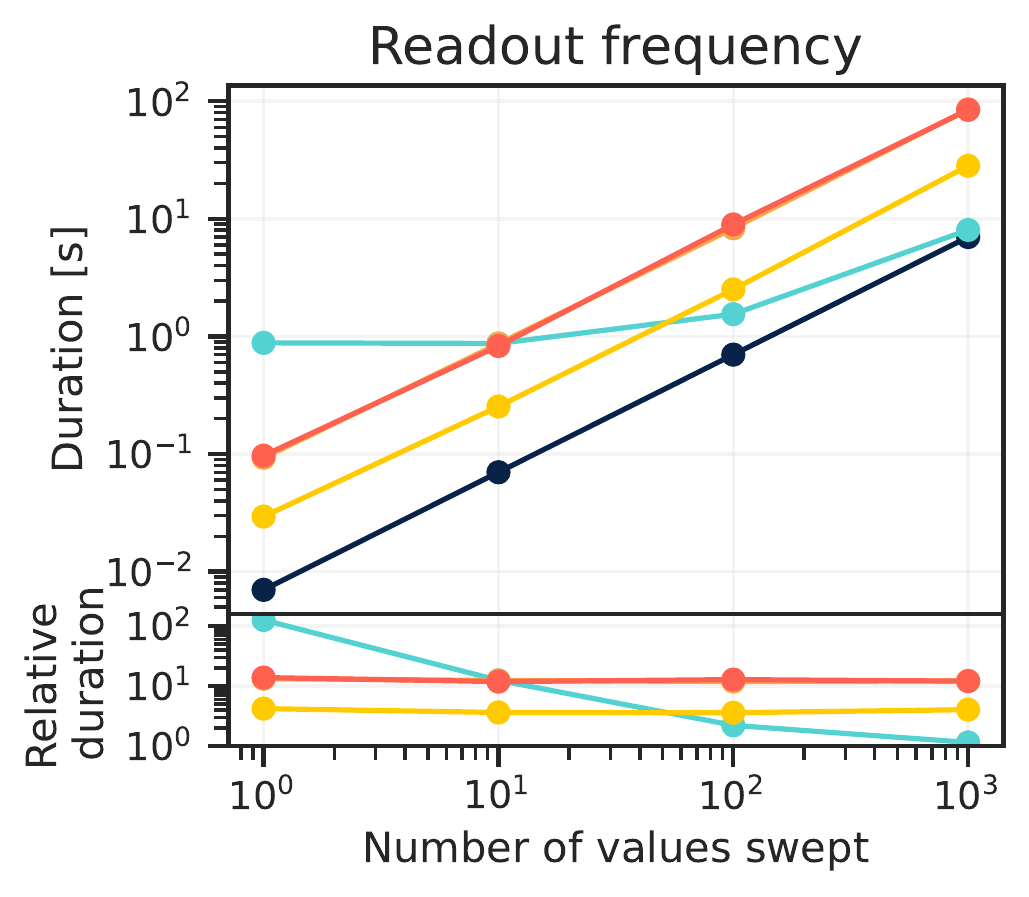}
		 &
		\includegraphics[width=0.33\textwidth]{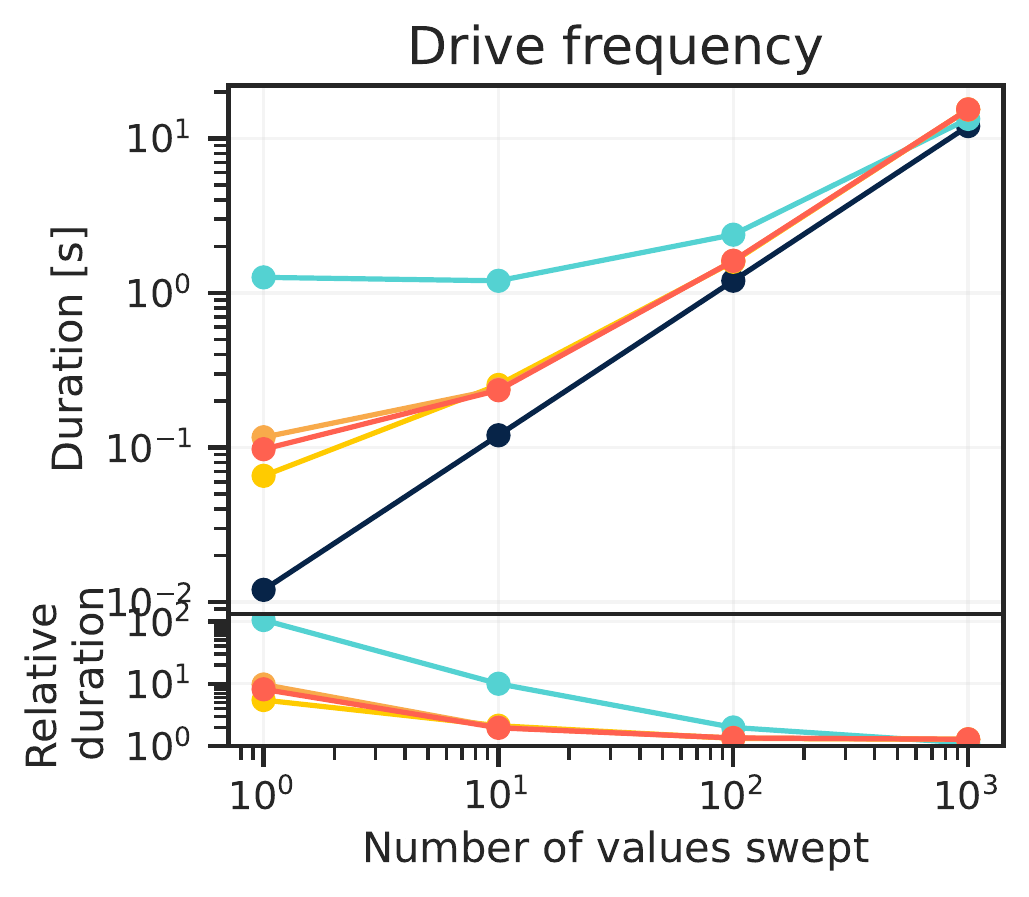}
		 &
		\includegraphics[width=0.33\textwidth]{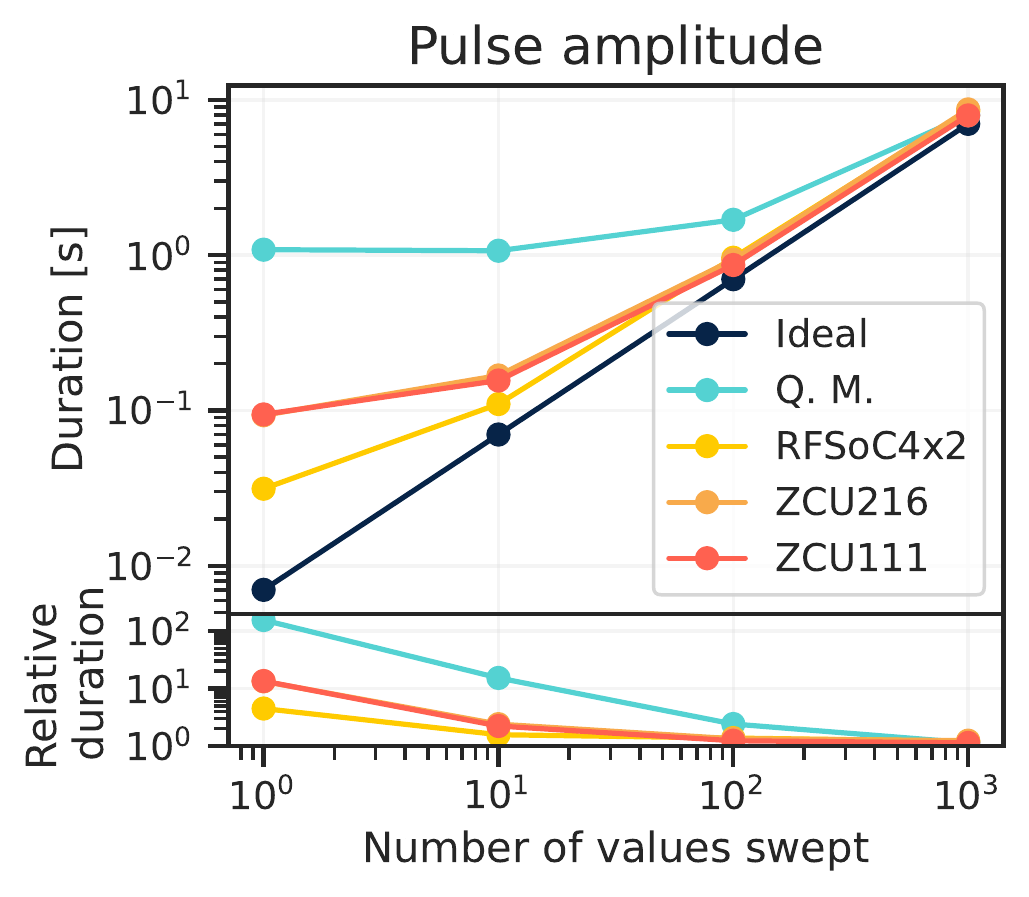}
		\\
		\includegraphics[width=0.33\textwidth]{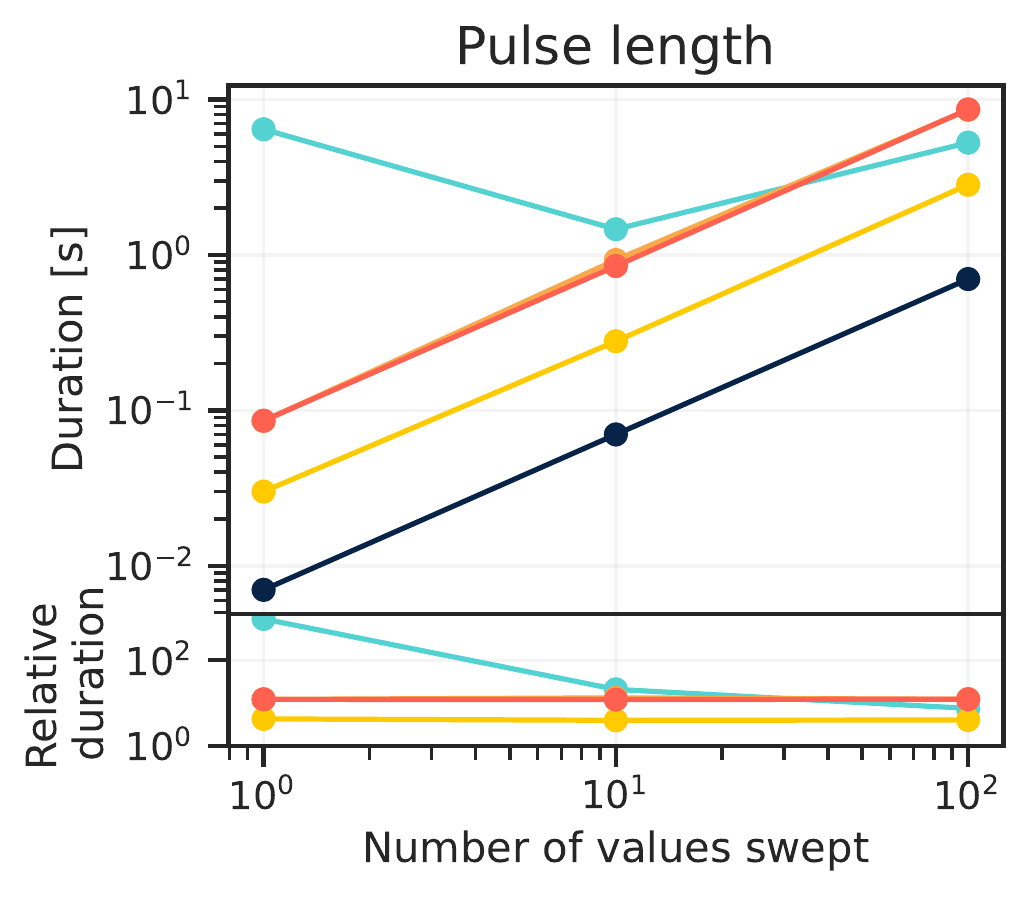}
		 &
		\includegraphics[width=0.33\textwidth]{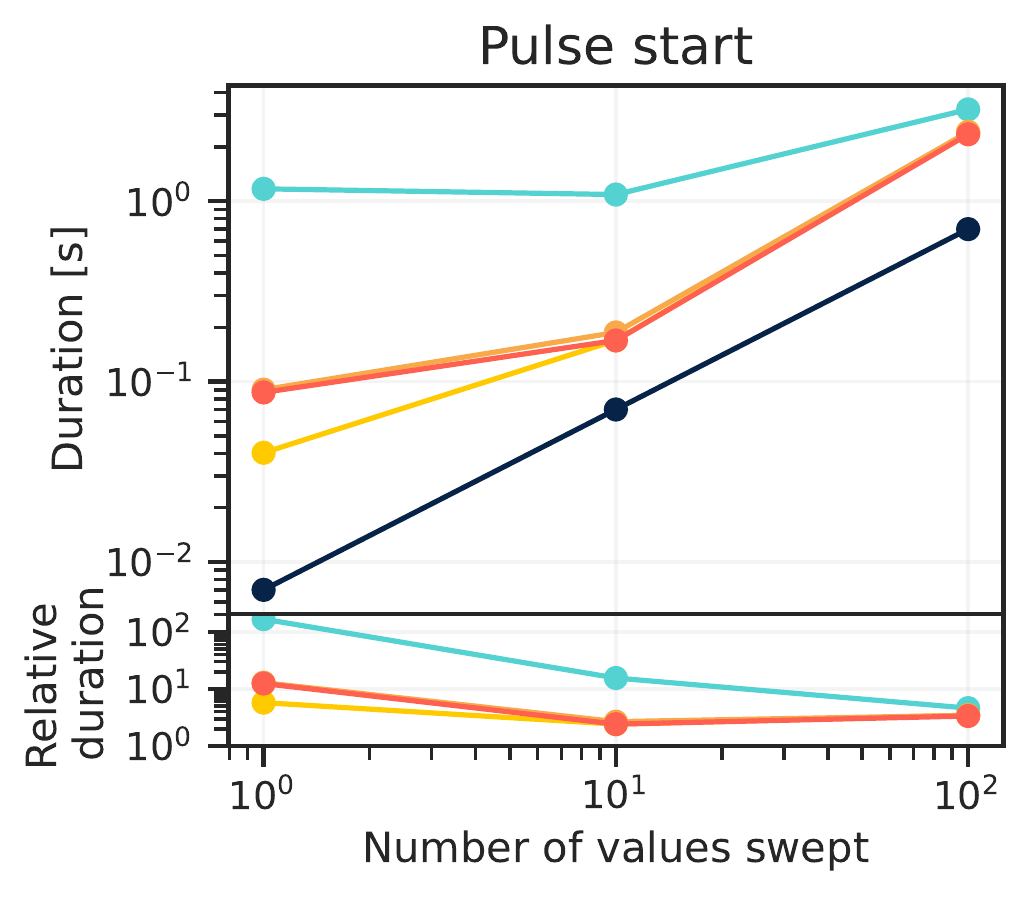}
		 &
		\includegraphics[width=0.33\textwidth]{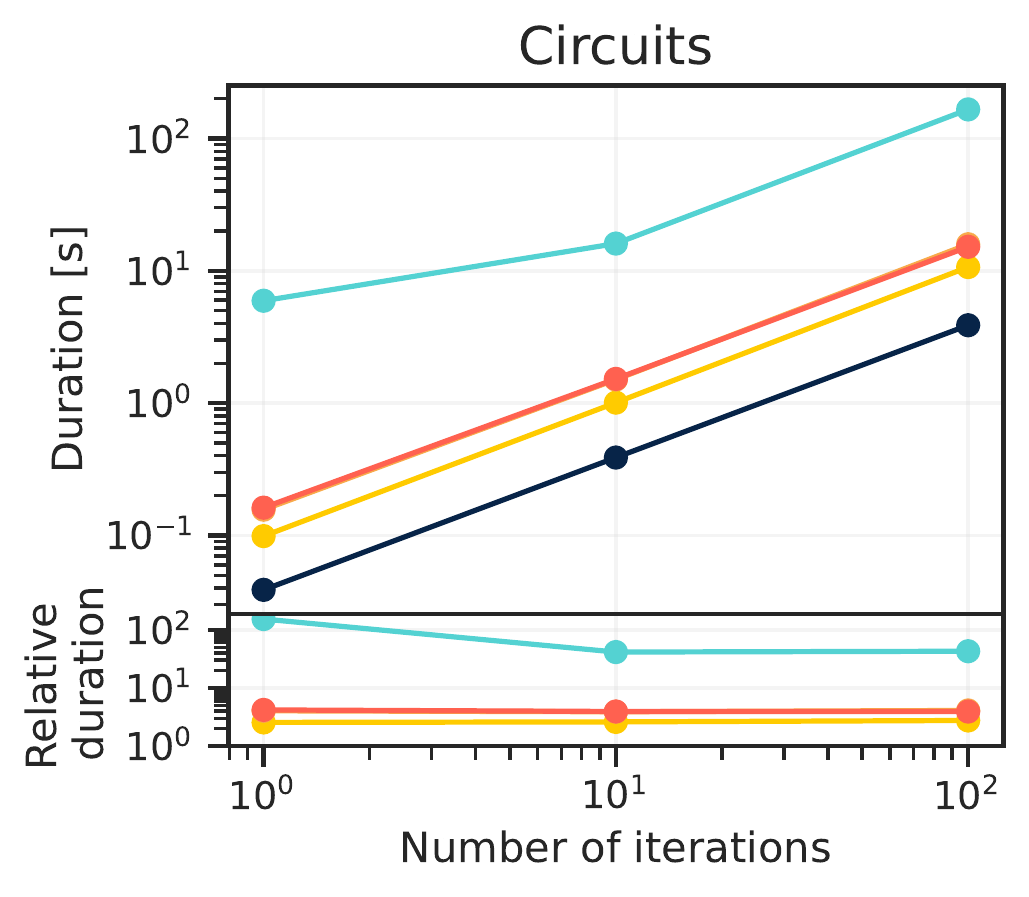}
	\end{tabular}
	\caption{The top plots illustrate the scaling of execution time with respect to the number of points in a sweep, while the bottom plots depict the ratio between actual execution times on various instruments and the minimum ideal time. ``Q.M.'' stands for Quantum Machines. The corresponding values were obtained using the same \Qibocal interface. The ideal time corresponds to how long the qubit is really used during the experiment, the difference from ideal and real time comes from software delays and communication latency.}
	\label{fig:benchmark_scaling}
\end{figure*}

We tested \Qibosoq using the \Qibolab API, benchmarking its speed on different RFSoCs and against commercial instruments (Quantum Machines, QBlox and Zurich Instruments).
The tested boards were the three Xilinx boards supported by \Qick, better detailed in \cref{tab:supported-boards}.
For better significance, in the plots only the results of Quantum Machines are shown to represent commercial instruments, as QM was the fastest among the three in the majority of cases~\cite{Efthymiou_2023}.

It is complex to do a proper and fair comparison between the software of different instruments,
while avoiding to compare the hardware. For this reason we decided to compare only speed of execution,
considering the deployment of experiment through the same interface, namely \Qibocal.

The results are presented in \cref{fig:speed_benchmark}.
The black bar in this plot provides the $\mathrm{ideal}$ time required for each
routine, calculated as
\begin{equation}
	\mathrm{ideal} =n_{\mathrm{shots}} \sum_i (T_{\mathrm{sequence}, i} + T_\mathrm{relaxation})
\end{equation}
where $T_{\mathrm{sequence}, i}$ is the duration of the whole pulse sequence
in the $i$-th point of the sweep, $T_{\mathrm{relaxation}}$ the time we wait
for the qubit to relax to its ground state between experiments,
$n_{\mathrm{shots}}$ the number of shots in each experiment and
the sum runs over all points in the sweep.
The $\mathrm{ideal}$ time denotes how long the qubit is really used during
an experiment and provides the baseline for our benchmark.

The experiments were repeated for $4096$ shots. Between shots, we waited a certain
amount of time: $5\,\mu$s for spectroscopies, $300\,\mu$s for the others.
A complete description of the experiments performed is given in~\cite[Appendix B]{Efthymiou_2023}.
Note also that the code required to reproduce these experiments is available at~\cite{qibosoq_runcards}.

Among the RFSoCs, we can see that the RFSoC4x2 consistently outperforms the ZCU111 and the ZCU216 in speed.
On the other hand, comparing it also with Quantum Machines, we can see that the \Qibosoq-controlled devices
are faster in six out of the eight performed experiments.

The key to explain the difference in speed between the \Qibosoq-controlled boards
and the commercial instruments can be found in the \textit{Ramsey detuned} experiment
timings. This routine involves sweeping the relative phase and the start time of a
pulse, a feature not supported by \Qibolab. Therefore, the experiment is performed
via the execution of various pulse sequences that get generated once at a time from
the client. This translates itself in a high number of communications between the
client and the device. In this regard, it is possible to note that the RFSoCs present
a much smaller overhead in the communication in respect to commercial instruments,
partially being explained with being composed of a single device and not with multiple
synchronized modules.
The communication overhead, including also any time required by the instrument to set
up, approximately corresponds to the first point of the plot presented in
\cref{fig:benchmark_scaling}, where the sequence execution time is negligible
with respect to the overhead itself.

The only experiment (among the one analyzed) where the RFSoCs are sensibly slower
than the commercial systems is the \textit{resonator spectroscopy experiment}
that can be conducted with a real-time sweeper, on hardware, for all the other
considered devices, while it's currently not supported by \Qibosoq and by the
standard \Qick firmware as shown in \cref{tab:supported-features}

In \cref{fig:benchmark_scaling}, the results of a benchmark on the scaling capability
of sweepers (increasing the number of points swept) and of circuits (increasing the
number of circuits executed) are presented.
For the readout and drive frequency, the experiment-template of resonator and qubit
spectroscopy was used, while for pulse amplitude and length
a Rabi-like experiment was performed. For the pulse start, a standard T1 experiment
was used and for the circuits a standard Randomized Benchmarking (RB)~\cite{Knill2008}.
The focus of this plot is, however, the parameter swept (or the number of circuits
executed) and not the experiments themselves.

We can clearly see which sweepers are not implemented with real-time sweepers
(on the readout frequency and on the pulse length), not considering the ``Circuits''
plot that cannot logically being implemented with sweepers.

For the majority of these plots we can see that, initially, where the effective
duration of the experiment is given almost completely by the overhead, the RFSoCs
perform much better than the commercial systems. Increasing the number of swept
values, however, decreases the difference of duration between instruments, with
the commercial ones that even becomes slightly better than
the RFSoCs for many-points scans.

\subsection{Calibration experiments}
\label{sec:calibration}

\begin{figure*}
	\begin{tabular}{ccc}
		\includegraphics[width=0.33\textwidth]{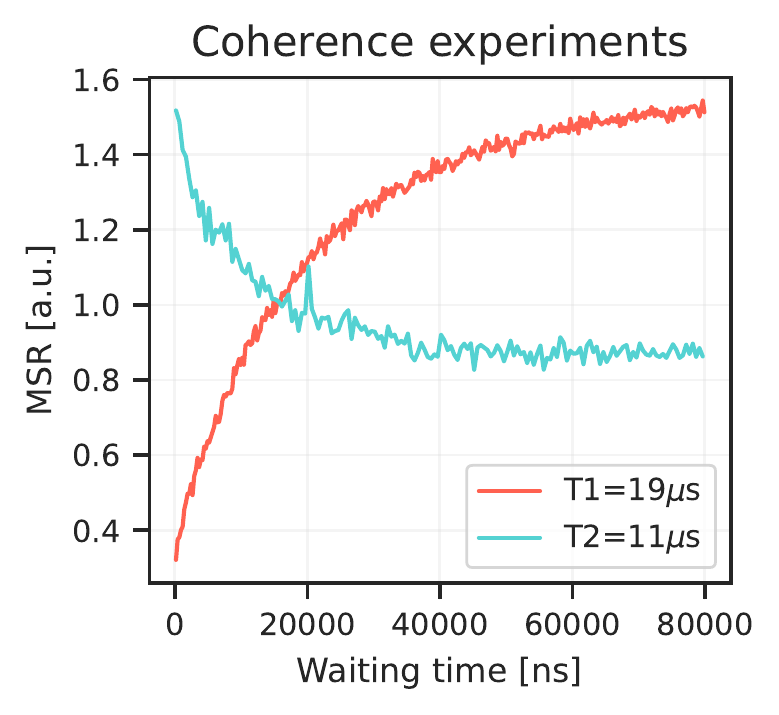}
		 &
		\includegraphics[width=0.33\textwidth]{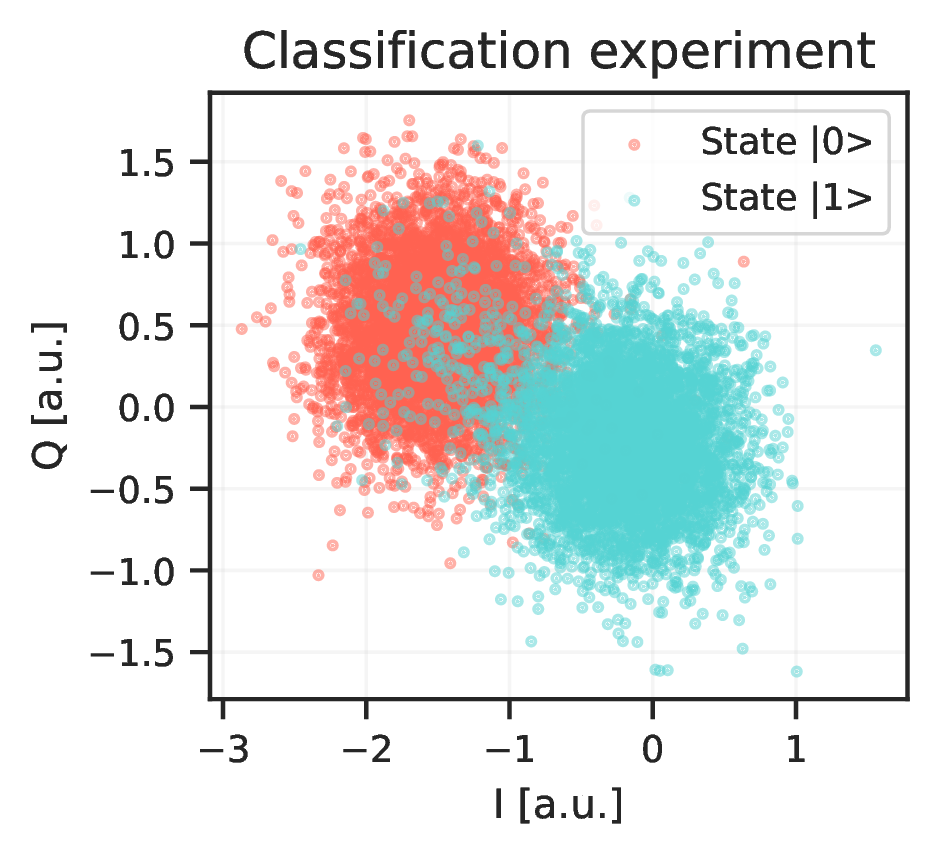}
		 &
		\includegraphics[width=0.34\textwidth]{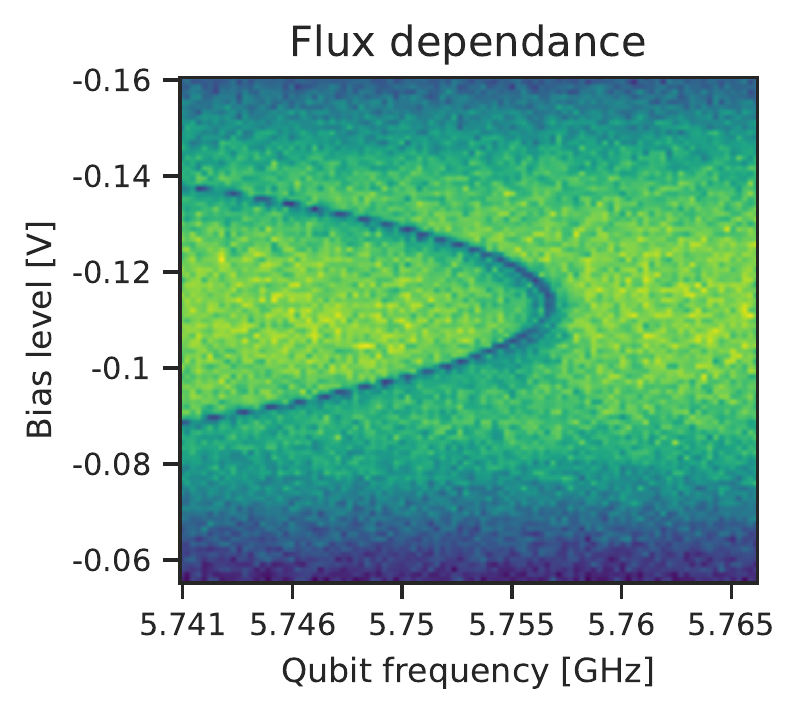}
	\end{tabular}
	\caption{Examples of calibration experiments. In the first experiment, the relaxation and dephasing time of a single qubit are computed. The second plot present a classification experiment from which an assignment fidelity of $0.95$ was computed. In the last experiment, the dependency of the 0-1 transition frequency to an applied bias was analyzed.}
	\label{fig:calibration_plot}
\end{figure*}

Alongside the presented speed benchmarks, we conducted tests of \Qibosoq on quantum
hardware, connecting the available boards to various single qubits, without flux
dependency, manufactured by TII (Technology Innovation Institute) and
a flux-tunable multi-qubit chip, by QuantWare~\cite{quantware}.

While performing a comprehensive quality benchmark is challenging due to the numerous
variables in calibration, the RFSoC-based system demonstrated competitiveness with
commercial instruments, reaching values compatible between instruments and in accord
with design parameters, considering multiple calibrations executed from scratch, both
for the assignment fidelity and the single-gate fidelity (computed with RB).

More generally speaking, there is evidence beyond this paper suggesting that RFSoC
hardware should perform at least on-par to standard hardware and that there is no
performance reason to prefer the latter in respect to the former~\cite{Kalfus2020,Gebauer2021,Tholen2023,Biznrov2024,Eriksson2024,Priyank2024}.
The differences therefore lie in speed and, in particular, software features and frameworks integration.

Some examples of the obtained results are shown in \cref{fig:calibration_plot}.

The initial two plots feature the Xilinx RFSoC4x2 connected to a single qubit within a 3D cavity.
In the first plot, we show the result of the $T_1$ and $T_2$ experiments.
For $T_1$, the relaxation time, we start each shot with the qubit prepared
in the excited state and then measure its value at different times. We expect
to see an exponential decay of which $T_1$ is the time factor.
For the $T_2$, we implemented the Ramsey experiment in the Echo version~\cite{Hu2018}.
We first prepare the qubit in an equal superposition of the two states via
a $\pi/2$ pulse, wait for a variable period of time $t$, apply a $\pi$-pulse
and than again a $\pi/2$ pulse after $t$. If the pulses are on resonance with
the qubit frequency, we again expect to see just an exponential curve, with
$T_2$ as time constant. The measured value are in alignment with design
parameters: $T_1 = 19 \mu$s and $T_2=11\mu$s.

As an example, we write here also a sketch of the \Qibosoq script to
execute the $T_1$ experiment (see the documentation for complete calibration
scripts~\cite{qibosoq_doc}):
\begin{lstlisting}
from qibosoq.components.pulses import (
  Gaussian,
  Rectangular
)
from qibosoq.components.base import (
  Config,
  Parameter,
  Sweeper
)

# We define the pi pulse
pulse_1 = Gaussian(
    frequency=1_000_000_000,  # pulse frequency in HZ
    amplitude=0.5,  # pulse amplitude in full scale units
    duration=500,  # pulse duration in ns
    type="drive",
    dac=0  # output channel
)
# We define the readout pulse
pulse_2 = Rectangular(
    frequency=1_500_000_000, amplitude=0.5,
    duration=2000, name="readout_pulse",
    type="readout", dac=1,
    delay_start=500, # delay from last pulse in ns
    adc=0  # input channel
)
sequence = [pulse_1, pulse_2]
# We define a sweeper on the time of the readout pulse
sweeper = Sweeper(
            parameters = [Parameter.DELAY],
            indexes = [1],  # index of the pulse to modify
            starts = [0.0],  # initial parameter value
            stops = [8.5],  # final parameter value
            expts = 1000  # number of steps
)
# We perform the experiment
server_commands = {
    "operation_code": OperationCode.EXECUTE_SWEEPS,
    "cfg": Config(),
    "sequence": sequence,
    "sweepers": [sweeper],
    "qubits": [qubit],
}
HOST = "192.168.0.0"  # Server IP address
PORT = 6000 # Default server port
i, q = execute(server_commands, HOST, PORT
\end{lstlisting}

If we wanted to leverage the already-written experiments in \Qibocal, then we would just need (after an initial configuration) to define a yaml \textit{runcard} file (see~\cite{andrea_pasquale_2023_7957542} for better reference):
\begin{lstlisting}
- id: t1
  priority: 0
  operation: t1
  parameters:
    delay_before_readout_start: 0
    delay_before_readout_end: 8500
    delay_before_readout_step: 10
\end{lstlisting}

This second method (leveraging \Qibocal) should be the preferred way of executing
already present since it enables the experimenter to avoid any code-writing, while
focusing only on physical parameters. Anyway, even the bare \Qibosoq code has
just information on control pulses and take care automatically of communication
and of any processor-level instructions.

The second plot shows the result of an assignment fidelity experiment, where the
qubit is first prepared in a state ($\ket 0$ or $\ket 1$ by applying a $\pi$-pulse
to excite the qubit) and then a dispersive measurement is performed, retrieving a
single IQ value. When plotting the result of multiple preparations and measurements
for the two states, we expect to see two well-discriminated areas.
In this case, we then discriminate the two states just by finding the straight line
that better divides the two groups. As per standard, we then use this method to assign every IQ value to one of the two states.
The number of states prepared as $\ket 0$ ($\ket 1$) and then assigned to the same state, divided by the total count of prepared states, renders the assignment fidelity.
In the plot presented, the resulting assignment fidelity observed was $0.95$.

The third plot involves a multi-flux-tunable-qubit setup controlled by the Xilinx
ZCU216 RFSoC. A two-tone spectroscopy was carried out as a function of a DC voltage produced by the RFSoC, which is linearly related to the applied flux bias to the
qubit SQUID that is used for tuning.
From this plot we can study the correlation between the 0-1 transition frequency and a bias current for the sample under test.

These three experiments are just examples of the various possibilities in which \Qibosoq
can be of use. These are standard characterization experiments that need to be executed
for every (eventually flux tunable) qubit calibration. Through \Qibocal, the user
would not need to write them from scratch, just focusing on experimental
parameters.
Moreover, they showcase different types of experiments: from left to right we have
parameter one-dimensional sweeps experiments, single-shot experiments and lastly an example of a two-dimensional sweep experiment.

\subsection{Circuit execution}

In this paper, we mentioned numerous times how \Qibosoq can be used to leverage
the \Qibo high-level API in order to directly deploy quantum circuit.
It is interesting and worthy to at least present an example of the interface a
user would use.

First of all we note that any circuit-based experiment require an initial
calibration, the output of which in this case has to be stored in a specific
configuration file employed by \Qibo in order to transcript a gate to a pulse.
In the following case, the calibration values (eventually produced with
\Qibocal), are used to generate a gate-based application:
\begin{lstlisting}
  from qibo import Circuit, gates

	NSHOTS = 5000  # number of repetitions
	# create a single qubit circuit
	circuit = Circuit(1)

	# attach Hadamard gate and a measurement
	circuit.add(gates.GPI2(0, phi=np.pi / 2))
	circuit.add(gates.M(0))

	# execute on quantum hardware
	qibo.set_backend("qibolab", platform="rfsoc")
	hardware_result = circuit(nshots=NSHOTS)

	# retrieve measured probabilities
	freq = hardware_result.frequencies()
	p0 = freq.get("0", default=0) / NSHOTS
	p1 = freq.get("1", default=0) / NSHOTS
\end{lstlisting}

We want to emphasise that although this application in itself is trivial, we are
showcasing an example of how a self-hosted (maybe even self-manufactured) qubit
can be used for algorithms without the need of any commercial instrument or software.

\Qibosoq has already be used in more interesting algorithmic application as
presented in~\cite{Efthymiou_2023}.

\section{Outlook}
\label{sec:outlook}

In this paper we have presented \Qibosoq, an open-source server-side software
for controlling RFSoC electronics via \Qick. This setup simplifies operation of
self-host quantum hardware platforms through \Qibo, a full-stack quantum
computing middleware framework.

We have outlined the present state of the project's structure, emphasizing the
significant features integrated into release \texttt{0.1.3}.
The software is at the stage of allowing applications
related to performance benchmarks through arbitrary pulse control and physics
experiments based on the quantum circuit representation respectively with the
APIs of \Qibolab and \Qibo.
Thanks to this integrations, \Qibosoq can be used to speed up characterization experiments as well as deployment of arbitrary gate-based algorithms. Therefore, it is an optimal solution for laboratories interested in developing the full stack of quantum computing (from the qubits to the algorithms).

As far as we are aware of, right now \Qibosoq is the only solution that allows
experimenters to control a superconducting qubit using exclusively open-source
software, while simultaneously enabling both low-level applications (like
characterization experiments) and high-level ones, with algorithms deployment.
Historically open-source and community-driven projects have been essential to the
development of standard computation, and because of this we believe that is quite
important to have a high-standard, open-source and full-stack quantum computing
framework.

In the future releases of \Qibosoq, we plan to extend its capabilities to
support multiple synchronized RFSoC boards and test the framework on other
quantum technologies such as trapped ions, neutral atoms and photonics among others.


The \Qibosoq module and all results can be reproduced using the code at:
\begin{center}
	\href{https://github.com/qiboteam/qibosoq}{\color{blue}\texttt{https://github.com/qiboteam/qibosoq}}.
\end{center}
while the complete package documentation, along with several examples, can be found at:
\begin{center}
	\href{https://qibo.science/qibosoq/stable}{\color{blue}\texttt{https://qibo.science/qibosoq/stable}}.
\end{center}

\acknowledgments This project is supported by TII's Quantum Research Center. The
authors thank the \Qick team for helpful discussion, comments on this manuscript
and support. S.C. thanks CERN TH hospitality during the elaboration of this
manuscript. A.G.~acknowledges support by the Horizon 2020 Marie
Sk\l{}odowska-Curie actions (H2020-MSCA-IF GA No.101027746).
This work also is supported by PNRR MUR projects PE0000023-NQSTI and CN00000013-ICSC.

 \bibliographystyle{apsrev4-2}
\bibliography{references}

\end{document}